\documentclass{article}

\usepackage{amsmath}
\usepackage{epsfig}
\usepackage{amssymb}
\usepackage{amsthm, amscd}
\usepackage{amsfonts}

\newtheorem{thm}{Theorem}[section]

\newtheorem{rem}[thm]{Remark}

\def\a{\alpha}

\def\e{\epsilon}

\def\l{\lq\lq}

\def\s{\sigma}

\def\tn{\textnormal}

\begin{document}

\title{Tilted Interferometry Realizes Universal Quantum Computation
in the Ising TQFT without Overpasses}

\author{Michael Freedman$^1$, Chetan Nayak$^{1,2}$,
and Kevin Walker$^1$\\
$^1$Microsoft Research, Project Q, Kohn Hall, University of California,
Santa Barbara, CA 93108\\
$^2$Department of Physics and Astronomy, University of California, Los Angeles, CA 90095-1547}

\maketitle

\begin{abstract}
We show how a universal gate set for topological
quantum computation in the Ising TQFT, the non-Abelian
sector of the putative effective field theory of the $\nu=5/2$
fractional quantum Hall state, can be implemented.
This implementation does not require overpasses or surgery,
unlike the construction of Bravyi and Kitaev, which we take
as a starting point. However, it requires measurements
of the topological charge around time-like loops
encircling moving quasiaparticles, which require the ability to
perform `tilted' interferometry
measurements.
\footnote{This manuscript has substantial
overlap with cond-mat/0512066
which contains more physics
and less emphasis on the topology.
The present manuscript is posted as a possibly useful
companion to the former.}
\end{abstract}

\newpage
\newpage

\section{Introduction}
In [BK] a universal set of gates $\{g_1, g_2, g_3\}$ for the Ising
TQFT, the non-abelian component of the Moore-Read state [MR]
proposed [GWW] for the $\nu=5/2$ fractional
quantum Hall (FQH) plateau [Wi,P,ECPW,X],
was constructed in an abstract context in which there
were no restrictions on the global topology of the space-time.
Clearly for a laboratory device the relevant space-time should embed
in $R^2 \times R^1$.  Presumably, simply adding
this constraint to the $BK$ context prevents the construction of a
complete gate set.  However if we add to their model a certain, we hope
realistic, assumption that the topological changes $1, \sigma$ and
$\psi$ can be distinguished on a simple (framed) loop $\gamma$
in space-time, then $\{g_1, g_2, g_3\}$ may be realized in $2+1$ dimensions.
Distinguishing, or more exactly, projecting to the charge sectors
$1, \sigma$, and $\psi$ according to the interferometry measurements
of [FNTW], as discussed in [DFN,BKS,HS],
resolves the identity into the sum of three
projectors: $1d=\widehat{1}\oplus
\widehat{\sigma}\oplus\widehat{\psi}$.  A further generalization is
however needed.  To realize gates $g_1$ and $g_2$ we need to measure
interference between paths $\gamma_1$ and $\gamma_2$ which cannot
simultaneously be projected into any (planar) space-time-slice.
What we propose is analogous to a $\lq\lq$twinkling'' double slit
experiment where the two slits rapidly open and close and though
never simultaneously open, produce an interference pattern.  We call
this $\lq\lq$tilted interferometry'' since the loop $\gamma =
\gamma_1 \cup \gamma_{2}^{-1}$ may have the property that it cannot
be deformed into any single time-slice and so must be tilted in
space-time.  We thank Ady Stern for pointing out that our $\lq\lq$tilted interferometry" in analogous to the second, electric Arharonov-Bohm effect [AB] where case $A_0$, must vary in time as the particle passes.  In our case the domain of the FQHE fluid will vary in time. We would be reluctant to assert that interferometry can
be performed along any knotted loop $\gamma$ but we need only fairly
simple $\gamma_s$.   To build the gates: $g_1$ and $g_2$ the link
along which we do interferometry has only one local max (min) per
component, i.e. is the $\lq\lq$plat of a pure braid.''  (The third gate,
$g_3$, is a simple braid generator and requires no discussion here.)
For reference:
\[
{g_1} = \left|
 \begin{array}{cc}
  1 & 0   \\
  0  & e^{\pi i/4} 
   \end{array}
\right|, 
{g_2} = \left|
 \begin{array}{cccc}
  1 & 0 & 0 &  0 \\
  0 & 1 & 0 &  0   \\
  0 & 0 & 1 & 0 \\
  0 & 0 & 0 & -1
 \end{array}
\right|, \textnormal{ and } {g_3} = \left|
 \begin{array}{cccc}
  1 & 0 & 0 &  -i \\
  0 & 1 &  -i & 0 \\
  0 & -i & 1 &  0   \\
   -i & 0 & 0 & 1
 \end{array}
\right|.
\]

\section{The possibilities for interferometry on the Ising TQFT}

In this section we briefly describe model experiments in the context
of the $\nu =5/2$ FQH state, some of which will be used to construct $g_1$ and
$g_2$ in section 3.  We begin without the time "tilt."  Consider a
disk of $\nu =5/2$ FQHE fluid in which current is injected at
$A$, withdrawn at $B$ and $C$.  Tunneling paths with amplitudes $t_1$ and $t_2$ are
marked.  An unknown topological charge resides on the antidot X.
\vskip.2in \epsfxsize=4in \centerline{\epsfbox{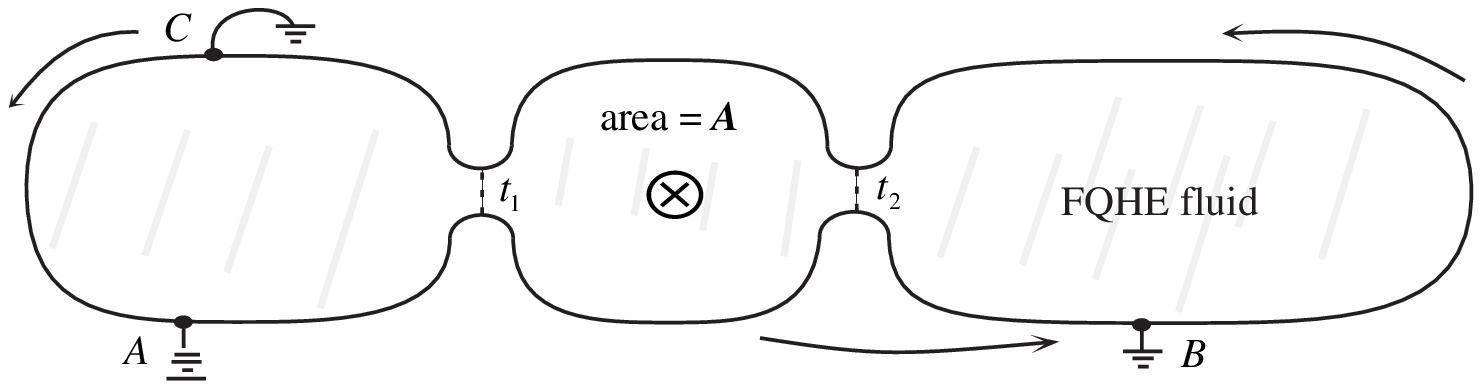}}
{\centerline{Figure 1a}}

\vskip.2in \epsfxsize=2in \centerline{\epsfbox{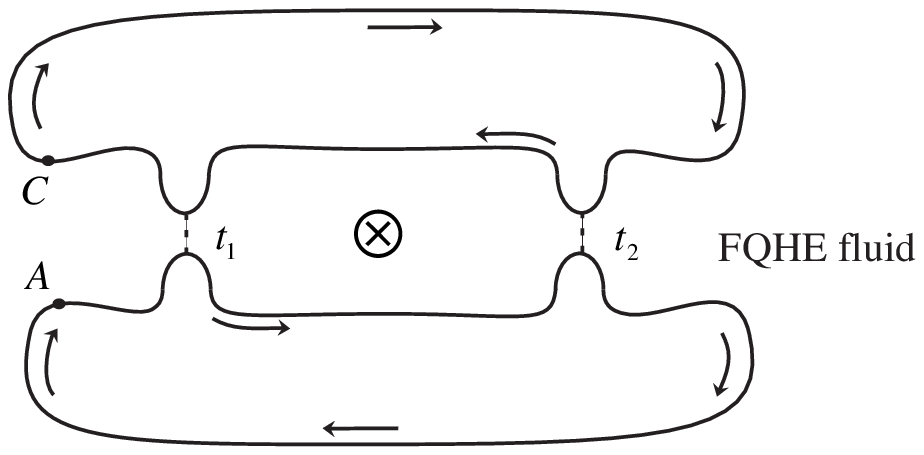}}
{\centerline{Figure 1b}}

Figure 1 shows two functionally equivalent setups.  In Figure 1a the FQHE $\lq\lq$fluid'' is on the $\lq\lq$inside'' and in the Figure 1b the $\lq\lq$outside'' of the bounding edge(s).


In space-time, $\lq\lq$braided tensor category,'' notation the two
tunneling paths for $\sigma$ particles contribute as:

\vskip.2in \epsfxsize=2.10in \centerline{\epsfbox{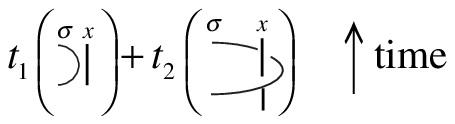}\label{fig:in2}}
{\centerline{Figure 2}}
(We ignore the $U(1)-$semion charges and the semi-classical
$B\cdot {\bf{A}}$ phase to concentrate our attention on the
more interesting non-abelian Ising charges:
 \vskip.01in \epsfxsize=1.35in \centerline{\epsfbox{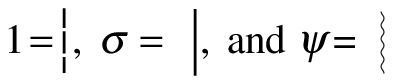}}
trivial, spin$ =1/2$, and spin$=1$.

Using the Kauffman rules:  \vskip.01in \epsfxsize=1.85in \centerline{\epsfbox{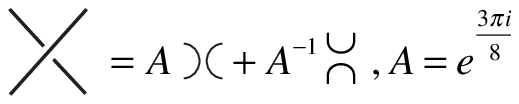}} which reproduce
the Ising rules up to the Frobeneous-Shur indicator (a sign which
arises in certain formulae but will not effect our results), and \includegraphics[width=.80in, height=.30in]{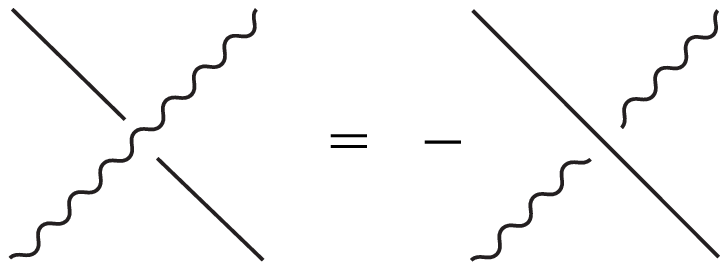} (Note: Using 
\epsfxsize=2.60in {\epsfbox{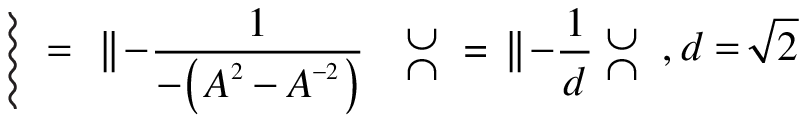}} the first rule implies the second.).  We evaluate the interference, Figure {~\ref{fig:in2}}, for $x=1, \sigma$, and $\psi$, and $t_1 = t_2$.

\begin{enumerate}
\item case $x=1$:
\vskip.2in \epsfxsize=3in \centerline{\epsfbox{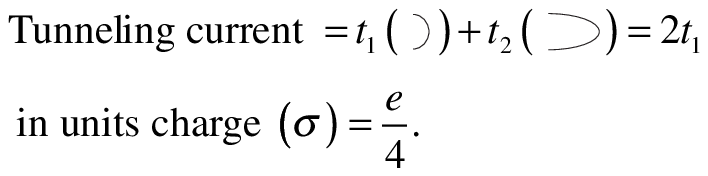}}

\item case $x= \sigma$:
\vskip.2in \epsfxsize=3in \centerline{\epsfbox{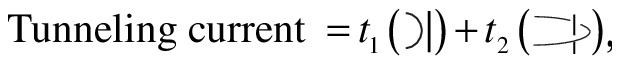}}
but if would be a mistake to algebraically combine the two processes since they represent orthogonal kets, which may be checked by pairing with external particles histories  \includegraphics[width=.40in, height=.20in]{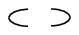} and \includegraphics[width=.45in, height=.25in]{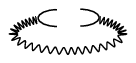}  . The results are \vskip.01in \epsfxsize=4in \centerline{\epsfbox{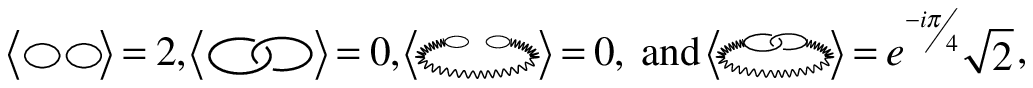}} as the reader will be able to check from the Kauffman rules (or the $S-$matrix - given later.) Orthogonality implies the norm of the combined processes is independent of the relative phase.  As observed in [HS,BKS], this orthogonality means no change in interference with changing area ${\bf{A}}$.  As ${\bf{A}}$ can be modulated with a side gate, this property should be experimentally accessible.
\item case $x=\psi$:
\vskip.2in \epsfxsize=3in \centerline{\epsfbox{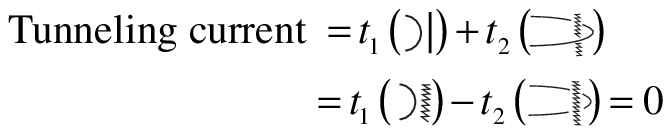}.}
\end{enumerate}

Formally these three outcomes for $x=1,\sigma , \psi$ are quite distinct.
Up to now have here only considered the $\sigma-$tunneling current.
One would also expect a smaller temperature dependent contribution
from $\psi-$tunneling which would have to be added to the
calculations above.  There would be terms:

\vskip.2in \epsfxsize=2.50in \centerline{\epsfbox{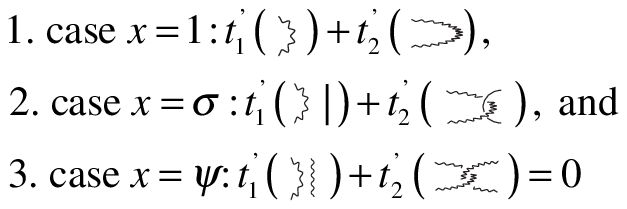}}
respectively.

In any case there is ample independence to expect a relatively simple\footnote{We do not belittle the experimental difficulties.  $\lq\lq$Simple'' is merely a comparison to what we will soon propose.}
interferometry measurements around $x$ to project into one of the
three sectors $1, \sigma$, or $\psi$.

If a simple loop $\gamma$ lies in a FQHE liquid at time $=t$ we may
project onto particle states $1, \sigma$, or $\psi$ along
$\gamma$ by an experiment which is a geometric distortion of, but topologically identical to, figure 1b.

{\vskip.2in \epsfxsize=3.5in \centerline{\epsfbox{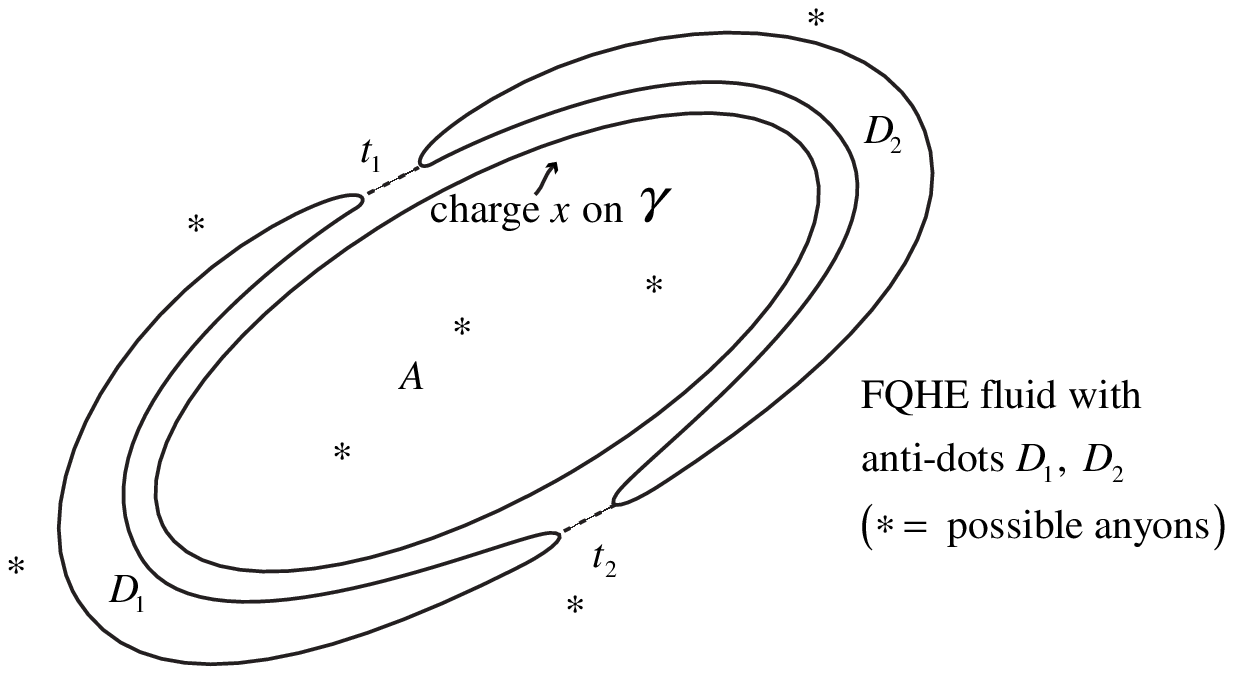}}
{\centerline{Figure 3}}}

The experiment suggested by Figure 3 is not tilted but
describable within a time slice.

Figure 3 depicts a plane filled with FQHE fluid except for two distorted anti-dots $D_1$ and $D_2$.  The asterisks represent quasi-particles. If we could measure the tunneling current between them (and vary area ${\bf{A}}$ as we do so), we project to a collective charge $1$, $\sigma$, or $\psi$ along $\gamma$.

Let us now take up tilted interferometry with $\sigma$ particles.
We expect technological limitations to
confine us to planar puddles of FQHE fluids at any times slice (i.e.
no $\lq\lq$overpasses'') and just as with MOSFET technology planarity
can be a sever constraint. But suppose a band of material (FQHE
fluid) $A$ is blocking a new band $B$ which we wish to construct,
might we break $A$, allow $B$ to pass, use $B$ for whatever purpose,
break $B$, and then reconstitute $A$?  If we could measure the
charge around the resulting time-like hole $\gamma$ in $A$ (See Figure 4.) and
if we found charge $=1$, it would be, as far as $SU(2)-$Chern-Simons theory
were concerned, as if $A$ were never broken.  This is the essential idea behind $\l$tilted interferometry''; in essence it is an $\l$end run'' around planarity constraints.

{\vskip.2in \epsfxsize=3.75in \centerline{\epsfbox{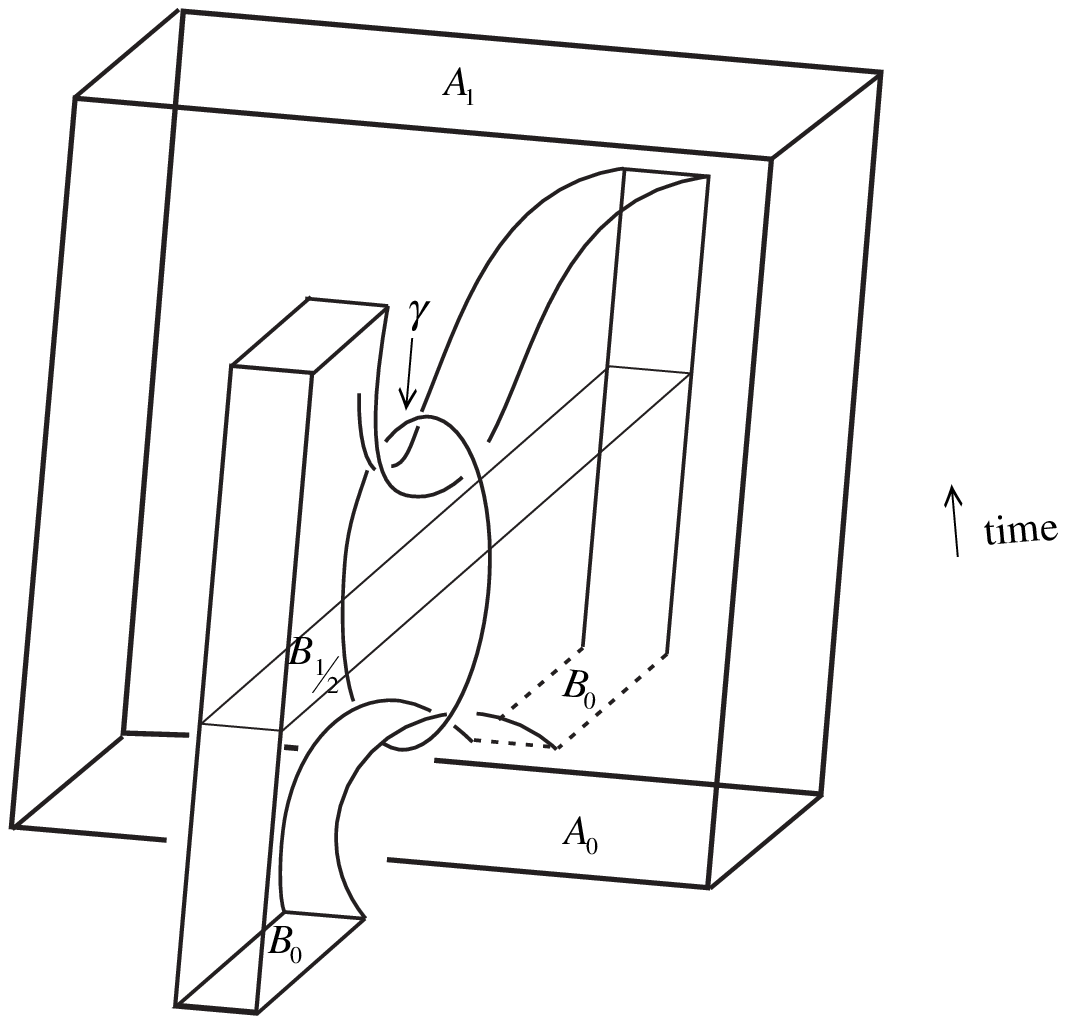}}
{\centerline{Figure 4}}\label{fig4}}

In this vein, consider the resistance between anti-dots $D_1$ and $D_2$ contained
in $A$ over a period of time in which $A$ is broken and rejoined.
If this time $A$ is broken is comparable to the tunneling time between $D_1$ and
$D_2$ (and various delays such as tortuous contours of the FQHE
fluid might be employed to achieve this) then the resistance should
depend on differences between the upper $\gamma_1$ and lower $\gamma_2$
tunneling trajectories as in Figure 5.  If we can prevent the environment from $\l$measuring'' charges on $D_i$, $i=1$ and $2$, $A$ may remain broken for longer.
{\vskip.2in \epsfxsize=3.75in \centerline{\epsfbox{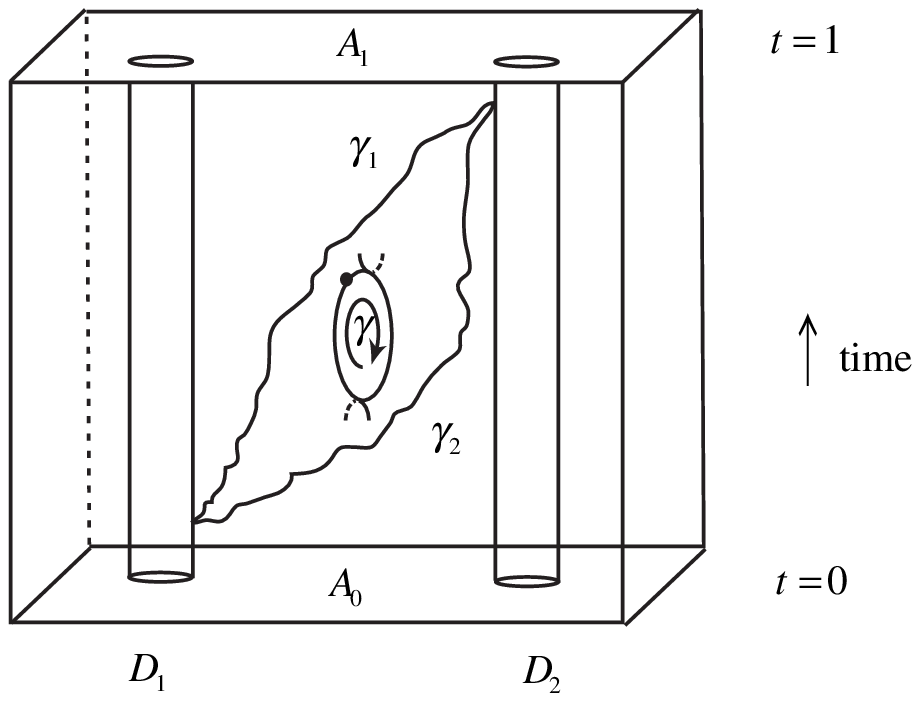}}
{\centerline{Figure 5}}\label{fig5}}
We suppose, here, that the experimental set-up is such that current is injected into $D_1$
near time $t_0$ and then withdrawn from $D_2$ near time $t_1$.

We now turn to the types of measurement needed to yield gates $g_1$ and $g_2$. 
In designing a gate, $\gamma$ might become complicated, needing to avoid some regions of space-time and pass through others.  In principle $\gamma$ might be a knotted in $(2+1)-$
space-time.  Fortunately, we only will need 
to measure the topological charge on a loop $\gamma$ with one max
and one min – in space-time (or a multi-loop where each component simultaneously shares this property).  

Note: For a simple loop on the
boundary of $(2+1)-$ space-time the projection into change super selection
sectors $\widehat{1} = \underset{\textnormal{charges } a}{\bigoplus}
\widehat{a}$ is mathematically well defined.  On the other hand if $\gamma$ lies in
the interior, a normal framing to $\gamma$ is required define this
define this decomposition (and different frames changes this
decomposition by more than phase factors as would be the case for
the $S-$matrix - conjugated decomposition).  In the $\l$untilted case'' the time arrow supplies a natural normal frame for the tunneling quasi-particle. In the tilted case, to produce a normal frame, a $\l$base-point'' needs to be defined on the anti--dots. One way to do this is to place a drain, like $\l B$'' in Fig 1a on each anti-dot.  This may or may not be difficult to arrange in a given implementation.  An alternative, described below, is to use one pair of anti-dots to define the boundary of interest and a second $\lq\lq$satellite'' pair for interferometry.  The relation of the satellite to the original defines the framing.

\vskip.29in \epsfxsize=1in \centerline{\epsfbox{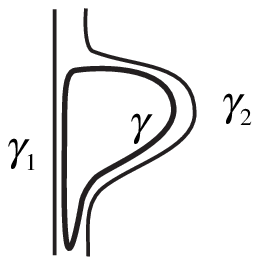}}
{\centerline{Figure 6a}}

\vskip.2in \epsfxsize=3in \centerline{\epsfbox{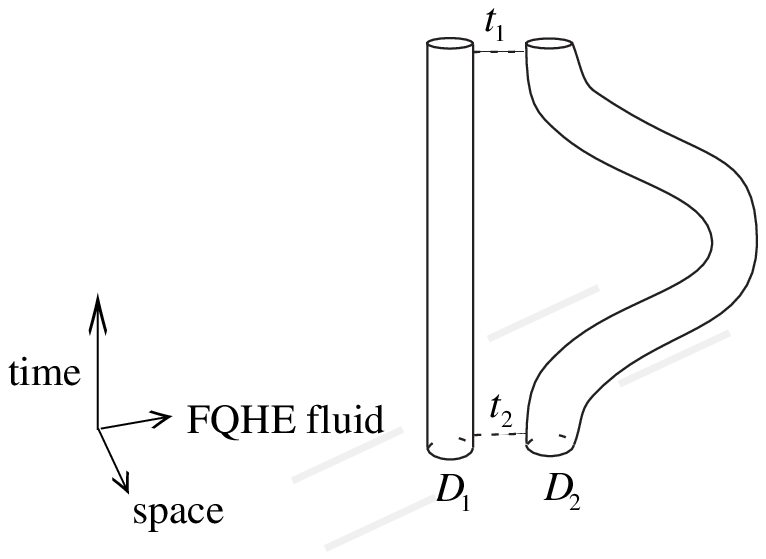}}
{\centerline{Figure 6b}}

Figure 6a-6b shows geometry for interferometry around a loop
$\gamma$ (with single space-time max and min) using one fixed
$\{D_1\}$ and one moving anti-dot $D_2$.

Let us consider the physical meaning of the framings
on the Polyakov loop is and how it
might be dictated.  First, how do we think of a tunneling particle?
As a Brownian path or a smooth arc?  In the former case it should be
impossible to assign a framing number, but energy considerations and
the finite size of quasi-particles suggest that 
most of the amplitude across a tunneling junction is concentrated on
the isotopy class of the obvious straight (and zero framed or $\l$time framed'') arc
across the junction.  We assume that the quasi particles do not
carry angular momentum while they tunnel.  A similar issue arises if
we transport a quasi-particle on a moving anti-dot.  How do we
control the rotation of the quasi-particle on the dot?  This
question is crucial, for without an answer, there is no distinction
between doing (titled) interferometry on $\zeta_i$ with framing $=-1$
$(\zeta_i, -1)$ and interferometry on $(\zeta_i , 0)$.  We will need to
control the framing of the space-time arcs along which we transport
antidotes.  A possible answer is to create an asymmetry: e.g. a
$\lq\lq$tear drop'' shaped anti-dot so that the edge has
an energy well to serve as a  natural base
point to record rotation.  Another possible answer is to create a drain (like $\l$B'' of Fig 1a) which a quasi-particle cannot pass.  However we follow a third approach.  We use one pair $D_1$ and $D_2$ of anti-dots to measure the charge on a loop lying in a bit of boundary defined by other anti-dots, e.g. $\overline{D}$, $D_3$ and $D_4$ in Fig 15.  Except for the direction of the time coordinate, this mimics the setup shown in Fig 1a if we regard the central oval as a charged boundary.

\begin{rem} \tn{Interferometry depends on maintaining a superposition
among possible tunneling events.  It will be challenging to avoid
$\lq\lq$measurement'' as the geometry and/or position of $D_2$ in the
FQHE fluid is changed, but we see no fundamental reason that this
should not be possible. We thank C. Marcus for his suggestion that a $\lq\lq$bucket brigade"
of anti-dots (particularly if charge drains are attached) may be easier to implement
than electro-statically moving an anti-dot.}
\end{rem}

\section{ Gate Protocols}

The qubits to be manipulated are spanned by the two fusion channels
in the Ising CFT:  \includegraphics[width=1.50in, height=.30in]{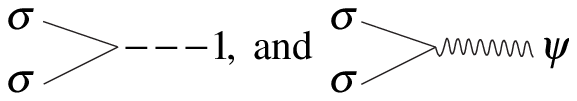}. 
Equivalently, this degree of freedom may be expressed in a single time slice: Consider a
twice punctured disk or $\lq\lq$pants'' $P$ (as part of a larger medium) in which the two
internal boundary components carry $\sigma$ and the outer boundary
carries $1$ or $\psi$ defining the basis of the qubit
$\mathcal{C}^2$.  See Figure 7.
\vskip.2in \epsfxsize=2in \centerline{\epsfbox{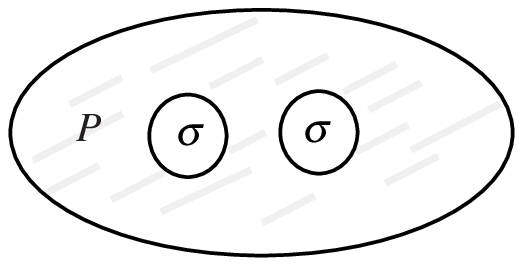}}
{\centerline{Figure 7}} The qubits can be represented as in
Figure 8.

\vskip.2in \epsfxsize=3in \centerline{\epsfbox{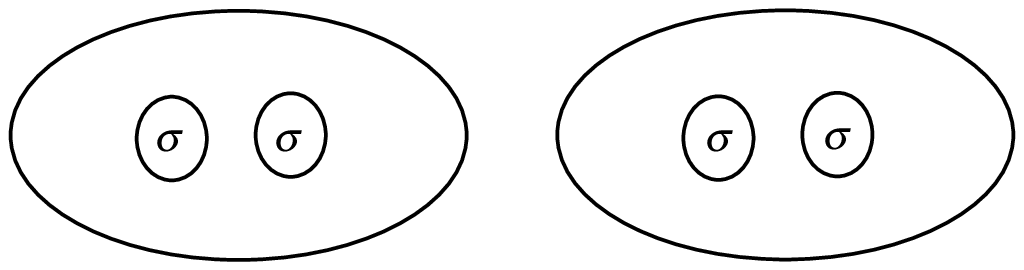}}
{\centerline{Figure 8}} Or more compactly as for Wilson (Abrikosov)
loop segments.

\vskip.2in \epsfxsize=2.5in \centerline{\epsfbox{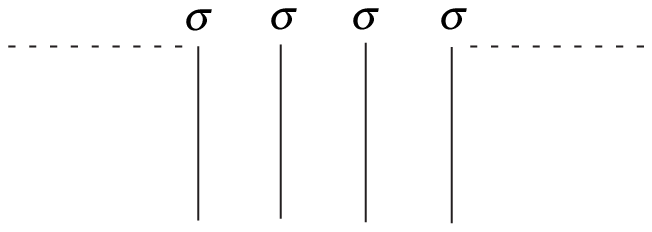}}
{\centerline{Figure 9}} 
It follows from the braiding rules of the
Ising TQFT (See [BK]) that
\vskip.2in \epsfxsize=3.50in \centerline{\epsfbox{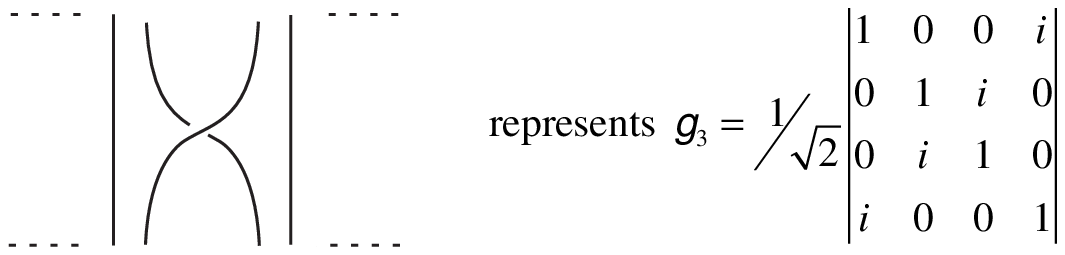}.}
{\centerline{Figure 10}} 
This gate requires no interferometry of any kind, it is simply a braid matrix.  Unfortunately the braid matrices in the Ising TQFT define discrete subgroups of $SU(N)$ so we are forced to use interferometry (or forbidden topology) to complete the gate set.  

The next gate we consider $ {g_2} =
\left|\begin{smallmatrix}
  1 & 0 & 0 &0 \\
  0 & 1 & 0 & 0 \\
  0 & 0& 1 & 0 \\
   0& 0 & 0 & -1
 \end{smallmatrix}\right|.
$
It is a $\lq\lq$controlled phase'' gate.  Since the $F-$matrix of the
Ising CFT is the Hadamar matrix: $ {\frac{1}{\sqrt{2}}} =
\left|\begin{smallmatrix}
1 & 1  \\
1 & -1
\end{smallmatrix}
\right| $

 which conjugates $\sigma_x$ into $\sigma_z$, producing $g_2$ is equivalently powerful to producing $\lq\lq$controlled NOT''
$ =  \left|
 \begin{smallmatrix}
  1 & 0 & 0 &0 \\
  0 & 1 & 0 & 0 \\
  0 & 0 & 0 & 1\\
   0& 0 & 1 & 0
 \end{smallmatrix}
\right|, $ but we follow [BK] in producing $g_2$.

\vskip.2in \epsfxsize=4in \centerline{\epsfbox{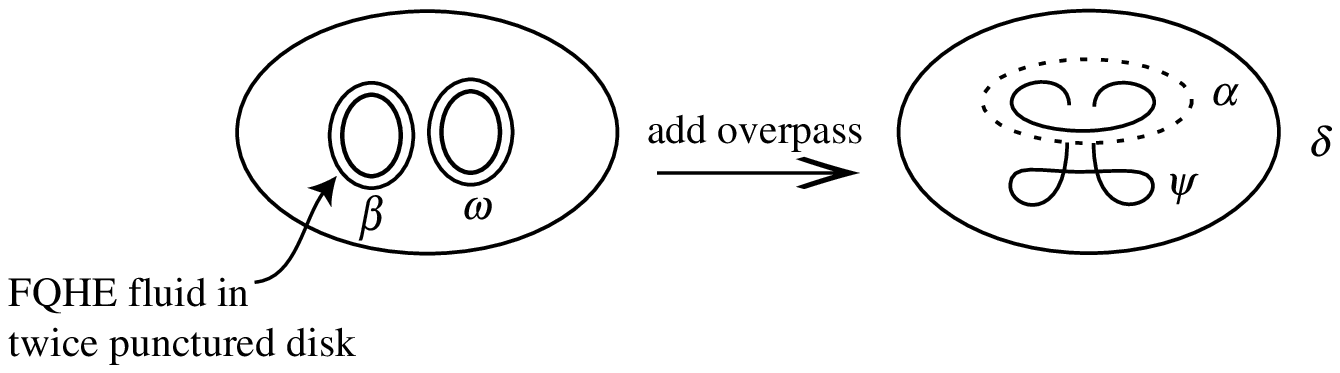}}
{\centerline{Figure 11}}

\vskip.2in \epsfxsize=3in \centerline{\epsfbox{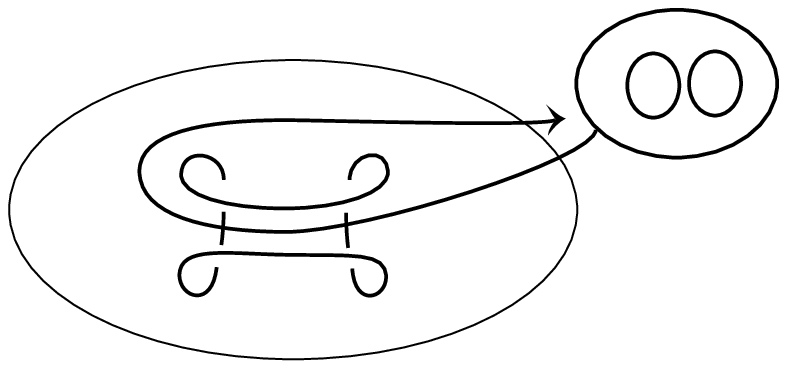}}
{\centerline{Figure 12}}

First we recapitulate in a geometric language the BK description which involves time slices
with \underline{$\tn{overpasses}$}, i.e. FQHE fluids which
cannot lie in the plane.  Then we will rearrange the time
coordinate and otherwise adjust the protocol so as we use only
planar fluids in each time-slice.  The price will be the need to use
$\lq\lq$tilted interferometry'' to project onto $SU(2)-$ charge
sectors along loops $\gamma$ which are titled in space-time.

The sum of the charges on the two resulting boundary components
$\delta$ and $\psi$ (Figure 11) is $1$ or $\psi$ according to
whether the charge along the overpass (dotted loop, $\alpha$) is
$1-\psi$ or $\sigma$.  (This follows from the $S-$matrices of
the theory:
\[
{S_{ij}^{0}} =   \left|
 \begin{array}{ccc}
  1/2 & \sqrt{2}/2 & 1/2 \\
  \sqrt{2}/2  & 0 & - \sqrt{2}/2  \\
  1/2 & - \sqrt{2}/2  & 1/2
 \end{array}
\right|
\]
 in basis: $1, \sigma, \psi$.  Furthermore $S^{\psi}_{\sigma, \sigma} = e^{i \pi/4}$ is the only nonzero entry for a punctured torus with boundary charge $=\psi$.)   

Ordinary, untilted, interferometry along $\psi$ projects into one of the states 
$1$ or $\psi$.  We hope to be in the sector charge $(\psi)=1$ and the probability
of this is $.5$, because $1$ and $\psi$ have equal quantum dimensions and therefore equal entropy.  If we are disappointed, we simply break the overpass and then reconstitute it.  Breaking the overpass returns the qubit to its original state.  This follows from a general principle (See Appendix A.) that adding quantum media is reversible simply by deleting what was added (whereas deleting quantum media is generally irreversible).    Reconstituting the band yields an independent $.5$ chance of getting the desired trivial charge on $\psi$.  We repeat as necessary until charge $(\psi) =1$ is observed.  Now charge $(\delta)$ and charge $(\a)$ are perfectly correlated; charge $(\alpha)=\sigma \Longleftrightarrow$ charge $(\delta)=\psi$ and charge $(\alpha) =1$ or $\psi \Longleftrightarrow$ charge $(\delta)=1$.

So far we have been manipulating the pants $P$ supporting the "control"
qubit.  Now take the $\lq\lq$controlled'' qubit and pass it, as a
body, around $\alpha$.  See Figure 12.

\vskip.2in \epsfxsize=3in \centerline{\epsfbox{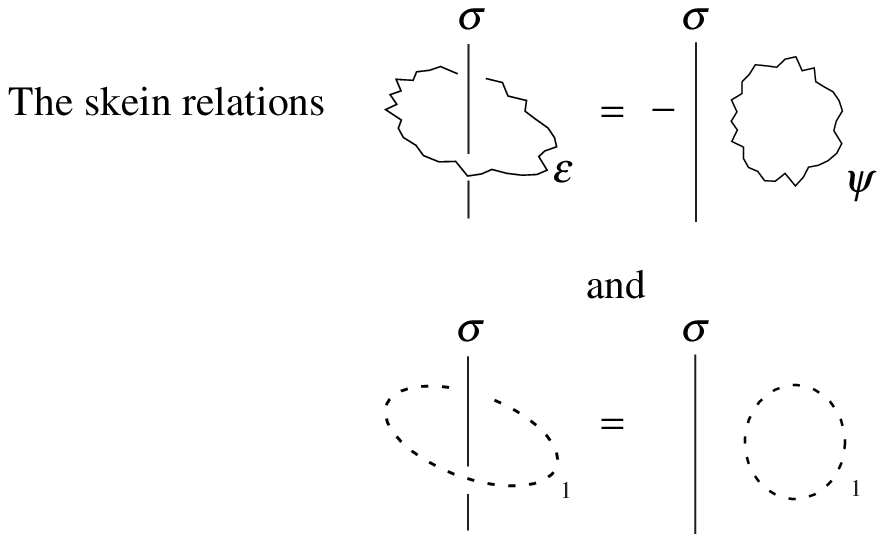}}
{\centerline{Figure 13}} tell us that the controlled qubit picks up
a phase of $-1$ if it is in state $|\psi\rangle$ and is unchanged if it is
in state $|1\rangle$.

Finally cut the overpass to return the two pants to there original
position.  The effects $g_2$. Figure 14 summarizes our
reorganization of $g_2$:
\vskip.2in \epsfxsize=5in \centerline{\epsfbox{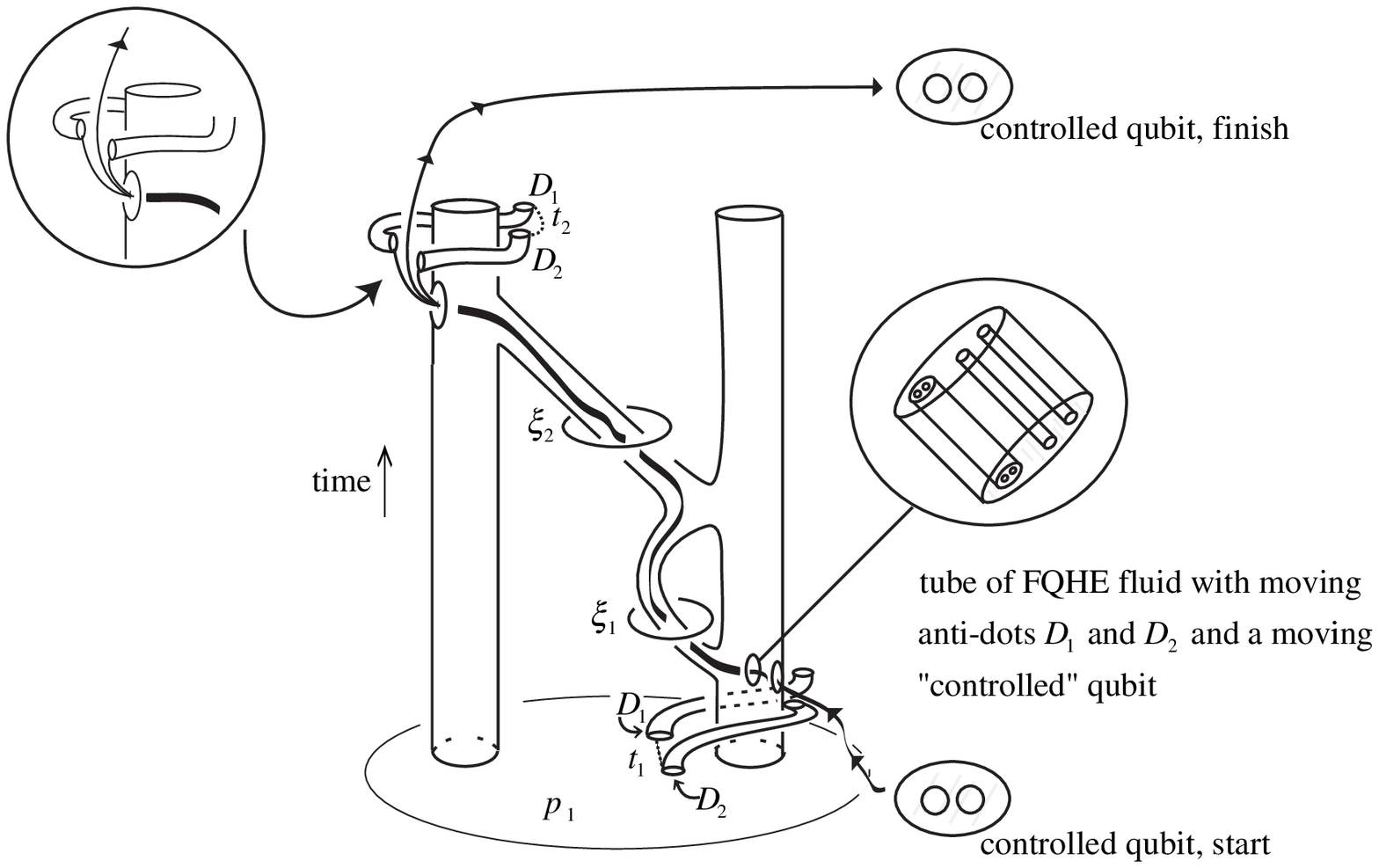}.}
{\centerline{Figure 14a}} For clarity Figure 14a is reproduced (expect
for the detour through $\xi_1$) in the slices in Figure 15.

To avoid clutter in Fig 14a we omitted an additional boundary component made from the space-time histories of the edges of an anti-dot $\overline{D}$ that divides into $D_3$ and $D_4$, which moves and later merges back into $\overline{D}$ as shown in Fig 14b.
\vskip.2in \epsfxsize=2in \centerline{\epsfbox{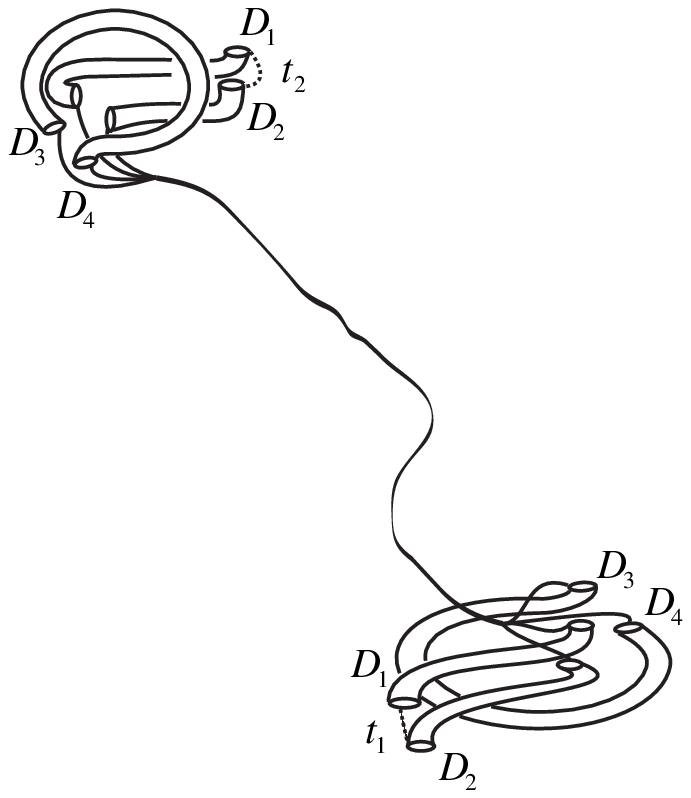}}
{\centerline{Figure 14b}}
\vskip.2in \epsfxsize=3.75in \centerline{\epsfbox{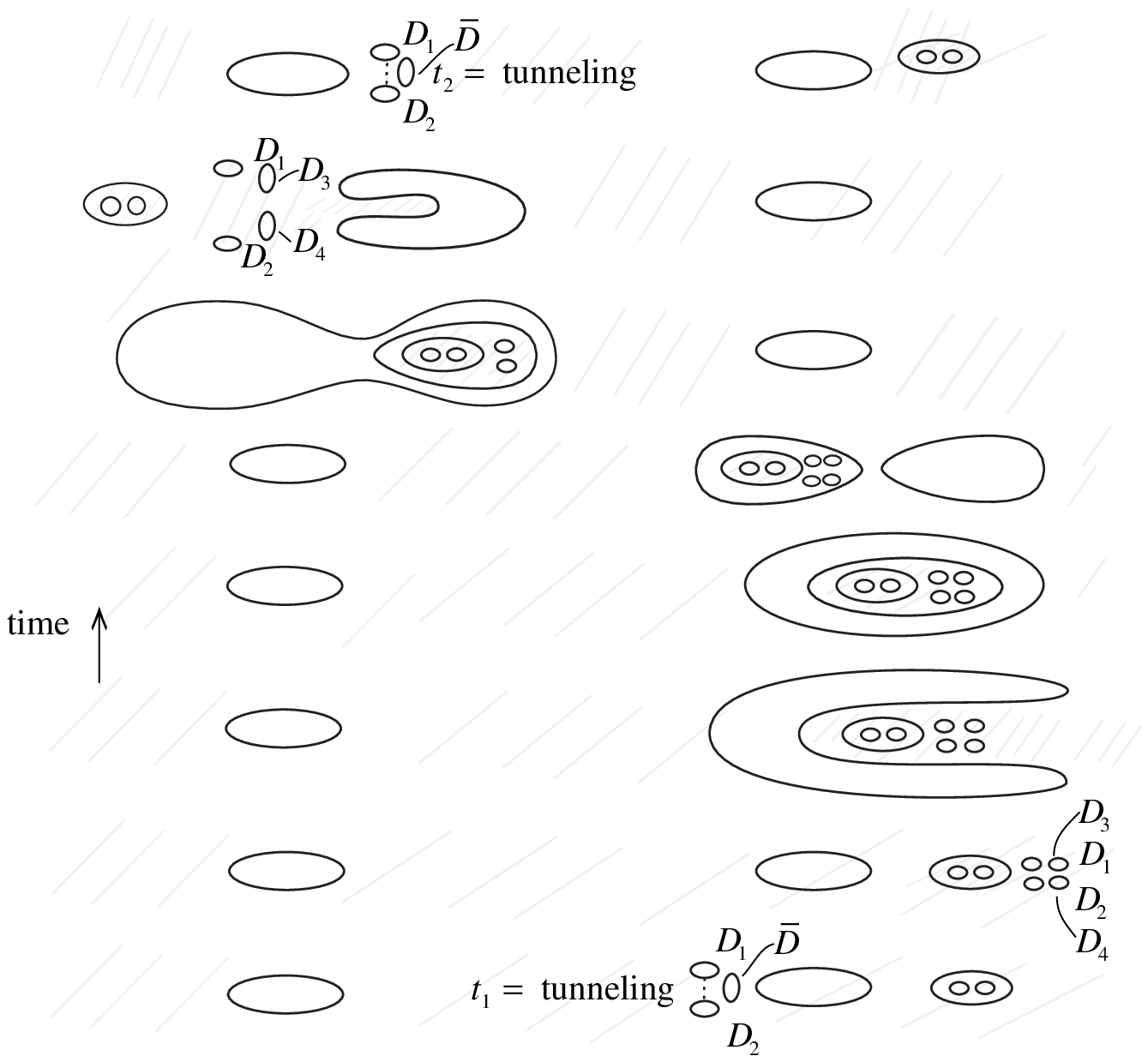}\label{fig14}}
{\centerline{Figure 15}}

In Figure 14a we do ordinary interferometry along the $\xi$ curves
and tilted interferometry along $\gamma =D_1\times$ time $\cup D_2
\times$ time $\cup t_1 \cup t_2$.  In words, we begin forming the
$\lq\lq$overpass'' band $B$, but now in space-time, and send the
controlled qubit, $q_2$, down the $\lq\lq$band'' $B$ as it is formed.  After a time,
the right puncture of the pants supporting the control qubit, $q_1$ splits and we
measure the $SU(2)-$ charge along $\xi_1$, hoping to observe $1$.
This would mean that the channel (Diagonal in Figure 14.) through which the band $B$ is
traveling does not disturb the structure of the first qubit $q_1$.  There are four
(equally likely if we neglect energetics associated to electric charge) topological charge splittings $\s {\nearrow 1\atop\searrow \s}$,
$\s {\nearrow \s\atop\searrow 1}$, $\s {\nearrow \e\atop\searrow
\s}$, $\textnormal{ and } \s {\nearrow \s\atop\searrow \e} $, so
there is a chance{\footnote{Charge energetics can be exploited to reduce to splittings: $\s {\nearrow 1\atop\searrow \s}$ and  $\s {\nearrow \e\atop\searrow \s}$, but interferometry is still required to distinguish these two (equally likely) cases.}} charge$(\xi_1)=1$.  If charge $(\xi_1)
\neq 1$ we fuse back (as shown) and try again until for some $i>0$,
charge $(\xi_i)=1$ ($i=2$ in figure 14).  When charge $(\xi_i)
=1$ we continue the tube across the pants supporting $q_1$ into the left puncture.  Terminate the band $B$ on the left side of the left puncture, allowing $q_2$ to complete its passage through the time-tilted overpass $B$.

First it is clear that the abortive attempts at building the band,
$\xi_1, \ldots \xi_{i-1}$, do not affect the qubit $q_1$ (except
possibly by an irrelevant overall phase): Splitting $a$ into $b
\otimes c$ and then re-fusing results in the original particle type,
{\includegraphics[width=.40in,height=.80cm]{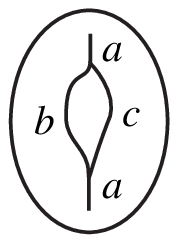}, is a multiple of the identity.  The
$\lq\lq$control'' qubit $q_1$ is clearly unaffected since the phase of the operator represented by the insert drawing is independent, by locality, of the state, $1$ or $\psi$, of the overall qubit $q_1$.

As before, $q_1$ will control a phase gate:$ \left|
 \begin{smallmatrix}
  1 & 0  \\
  0 & -1
 \end{smallmatrix}
\right| $  on $q_2$ iff the charge measure along $\psi$ is $1$;  $\psi$ in
Figure 14 is the difference of the two tunneling paths, $t_1$ and
$t_2$ between the moving anti-dots $D_1$ and $D_2$; it is the analogue of $\psi$ in Figure 11.  If $\psi$
is, instead, measured along $\psi$, then the gate has inadvertently interchanged the roles of $1$ and $\psi$ within the controlling qubit $q_{1}$;  a short calculation shows that $ \left|
 \begin{smallmatrix}
  1 & 0 & 0 & 0 \\
  0 & -1 & 0 & 0 \\
  0 & 0 & 1 & 0 \\
  0 & 0 & 0 & 1
 \end{smallmatrix}
\right| $ has instead been affected.  This is not too serious since
repeated application of the protocol gives a random walk in the
group $Z_2 \bigoplus Z_2$ generated by
 $
\left|
 \begin{smallmatrix}
  1 & 0 & 0 & 0 \\
  0 & 1 & 0 & 0 \\
  0 & 0 & 1 & 0 \\
  0 & 0 & 0 & -1
 \end{smallmatrix}
\right| $ \tn{ and }
 $
\left|
 \begin{smallmatrix}
   1 & 0 & 0 & 0 \\
  0 & -1 & 0 & 0 \\
  0 & 0 & 1 & 0 \\
  0 & 0 & 0 & 1
 \end{smallmatrix}
\right| $.  Our $\psi-$measurements tell us where we are within $Z_2
\bigoplus Z_2$ as we randomly walk; we simply halt upon reaching $ \left|
 \begin{smallmatrix}
  1 & 0 & 0 & 0 \\
  0 & 1 & 0 & 0 \\
  0 & 0 & 1 & 0 \\
  0 & 0 & 0 & -1
 \end{smallmatrix}
\right|. $  The tails on $\lq\lq$long walk'' decay exponentially so
this delay is acceptable.

Perhaps more serious is the fact that the anti-dots $D_1, \ldots, D_4$ must
be threaded, along with $q_2$, through the band $B$.  $D_1, \ldots,  D_4$
should be kept outside tunneling range and the two $\sigma$ charges
inside the pants $P_2$ carrying $q_2$ must not be fused.  This implies some geometric
constraints.  Clearly the size of the pants $P_1$ supporting $q_1$
must be enlarged, relative to the pants $P_2$ supporting $q_2$
before $q_1$ can  be used to control the phase of $q_2$.  This will
be only one of many technological challenges.

With this example of a gate implementation in hand, it makes sense
to discuss the general strategy and fundamental principles involved.
The general $3-$manifold $M$ with boundary does not imbed in $R^3 =
R^2 \times R$ but after puncturing $M$ by removing a collection of
proper arcs $M^\prime = M  \diagdown$ arcs will imbed.  If the
linking circle to a puncturing arc is measured to have charge $=1$
then the puncture is irrelevant (at least within an 
$SU(2)-$Chern-Simons theory).  Thus the strategy is to find some
protocol of puncturing and measuring which reduces the topologically
intricate gates of [BK] to sequence of planar time-slices.

\begin{rem}\label{3.0} \tn{ In calculating  (See [BK] for algorithms) the $SU(2)-$CS partition function $Z$ for the space-time history of a puddle of $\nu=5/2$ FQHE fluid it is only the intrinsic topology of the resulting $3-$manifold which is relevant and not imbedding in $R^3$.  We give some example to clarify this important point.  Our computations intentionally ignore linking not detectable within the space-time.}
\end{rem}
Suppose a pair of $\sigma$'s is pulled out of the vacuum and then
fused.  They will annihilate.  If however after there births the
fluid is cut to separate them, all correlation is lost.  If the
fluid is then rejoined and the $\sigma$'s fused the results will be
$1$ with probability $1/2$ and $\e$ with probability $1/2$.  See
Figure 16.
\vskip.2in \epsfxsize=4.5in \centerline{\epsfbox{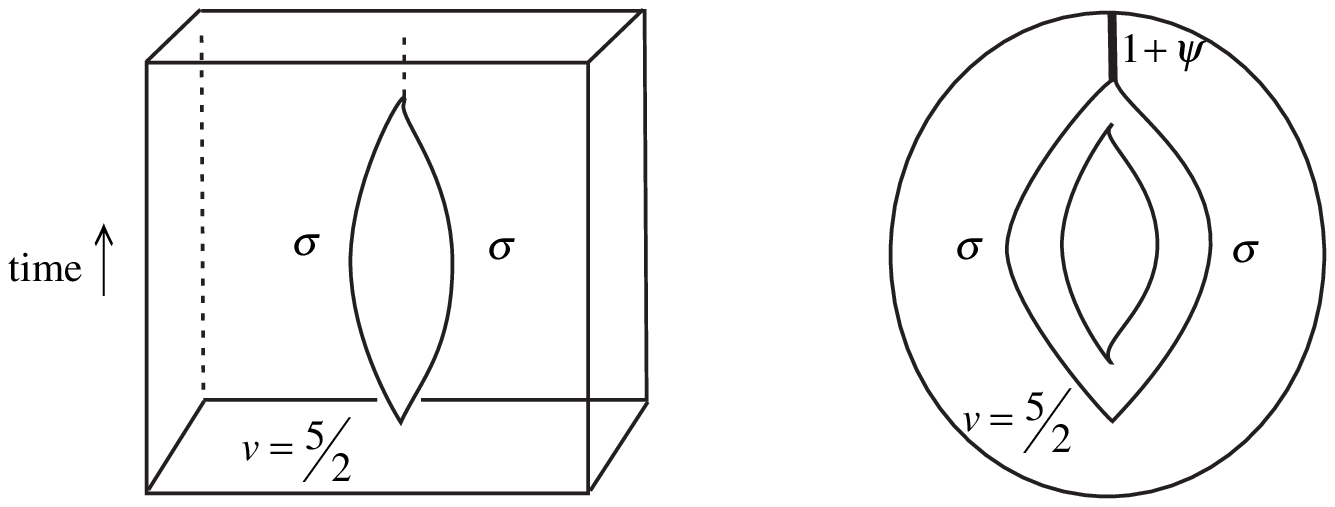}.}
{\centerline{Figure 16}}

Similarly since the $S-$matrix entry $S^ {1} _ {\sigma \sigma} =0$
two simply linked $\sigma$ trajectories cannot occur in a box of
space-time fluid, but can occur if only part of the box is filled
with $5/2$'s fluid.  See Figure 17.
\vskip.2in \epsfxsize=5in \centerline{\epsfbox{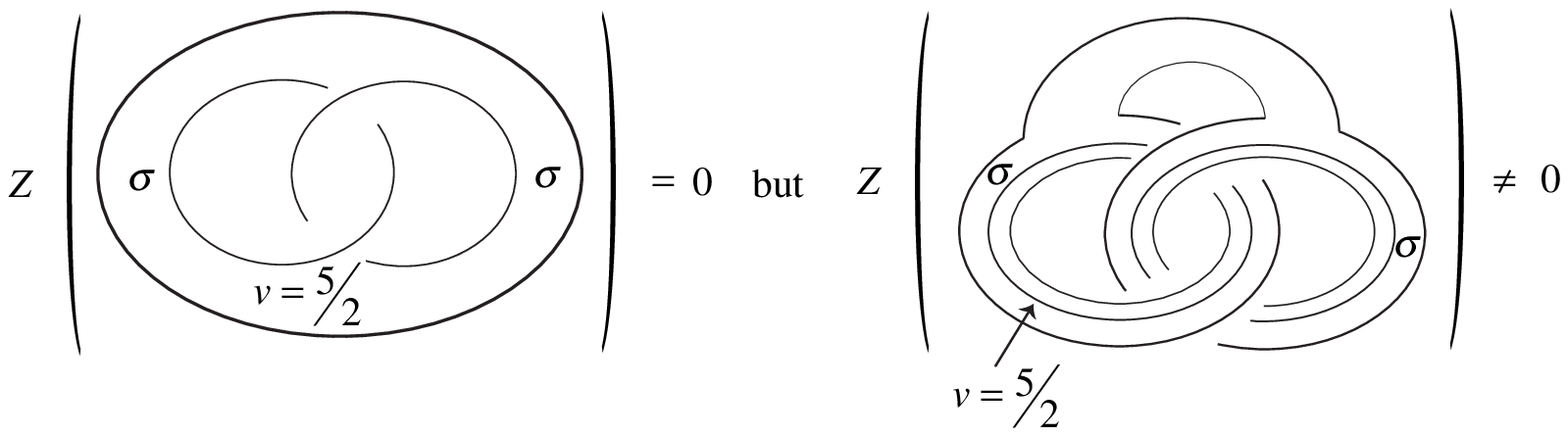}.}
{\centerline{Figure 17}}

\begin{rem} \tn{ The proceeding remark applies equally to the effective $U(1)$ and $SU (2)-$sectors of the theory.  We have been suppressing the fact that CFT modeling $\nu=5/2$ is a semi direct product of the Ising CFT (a variant of $SU(2)-$level  $=2$) and a $U(1)-$semionic theory $(U(1)-$level $=2)$ in order to concentrate on the more interesting nonabelian charges.  Certainly outside the FQHE space-time there is no sensible $SU(2)-$connection (or an effective $U(1)$ potential), which could mediate topological interaction, hence Remark \ref{3.0}.  On the other hand, the $U(1)$ gauge potential of ordinary  electromagnetism pervades all of space-time and it will produce Arharonov-Bohm interactions without regarded to the boundaries of FQHE fluid.  However, Since the particles in this theory carry electric but not magnetic charges the $U(1)-$ corrections are proportional to flux $B$ through the relevant surface (See [DFN].) and are easily made.}
\end{rem}
We turn no to the final gate $ {g_1} =  \left|
 \begin{smallmatrix}
  1 & 0 & \, \\
  0 & e^{ \pi i/4}
 \end{smallmatrix}
\right|. $ The description in [BK] may be summarized as:

\begin{enumerate}
\item Beginning with a qubit $q$ on a pants $P$, attach a tube to $P$ to obtain a punctured torus.  This is done by first adding a band $B$ and then measuring a charge $1$ or $\psi$ on the new boundary component.  If $1$ is measured the tube is regarded as successfully attached; if $\psi$ is measured then break the band and try again.
\item Let $D$ be Dehn twist in the curve labeled $\alpha$ in Figure 11. Act on $T$ by $D^2$.
\item Cut the band $B$ to change $T$ back into $P$. 
\end{enumerate}

Steps $1,2,3$ effect $g_1$. The computation follows from knowing the $S-$matrices and
twist{\footnote{In the Ising TQFT $\theta_1 =1$, $\theta_\s = \e^{2\pi i/16}$, and $\theta_\e =-1$.}}  parameters  $\theta$.  If $q$ is in state $| 1 \rangle$ the
charge along $\alpha$ is $1+ \psi$,  $\theta_1 =1$ and
$\theta_\psi = -1$, so under $D^2$, $(\theta_1)^2 =(\theta_\psi
)^2 =1$ is applied and no phase change occurs.  On the other hand, if $q$ is in state
$| \psi \rangle$ then the charge along $\alpha$ is $\sigma$ so
$D^2$ changes phase by $(\theta_\sigma)^2 = (e^{i \pi/8})^2= e^{\pi
i/4}$.

{\bf{Note:}}  If $D$ were used instead if $D^2$ the result would not
operate on the qubit since the charge on the internal punctures
would not return to $\sigma$ after the band $B$ is cut: it would be
$\frac{1}{\sqrt{2}}(1+\psi)$.

Our proposed implementation of $g_1$ closely follows this $3-$step
description.  To understand the protocol, refer back to Figure 14a.
Most of the that figure depicts activity on $P$ with the second
pants $P_2$ being threaded through a passage roughly from southeast
to northwest.  Since $g_1$ is a $1-$qubit gate, we dispense entirely with $P_2$; instead we thread two unlinked loops, 
with framing $-1$, $\zeta_1$ and $\zeta_2$ through this channel as
sketch Figure 18 (to be $\lq\lq$overlain'' on Figure 14a).  We will
need to use tilted interferometry to measure the charge on $\psi$, $\zeta_1$, and $\zeta_2$.

Thus each $\zeta_i$, $i=1,2,$ consists of a moving anti-dots $D_{i}^{\pm}$ carrying a $\pm \frac{e}{4} -\s-$ particle accompanied by companion anti-dots $D^\prime_i$ and $D^{\prime \prime}_i$ (moving or in $\l$bucket-brigade'') determining a framing of $\zeta_i$ $(\zeta_i, -1)$ by the normal direction from $D_{i}^{\pm}$ to the satellite anti-dots..  The role of $D_{i}^{\pm}$ is to carry a meridional $\s-$charge while tunneling $|t_1 -t_2|$ is measured between $D^\prime_i$ and $D^{\prime\prime}_i$.

We have added a new feature, we have assume that in preparing the anti-dots $D_{i}^{\pm}$, that we can pull out of the vacuum and later annihilate $\pm$ pairs of $\s$s.  The reason for this constraint is to restrict to two cases $1$ or $\e$, the possible outcomes of each$(\zeta_i , -1)$ measurement.  Indeed, the calculation for the change of basis from the meridial basis (in which $\s$ would surely be measured) to the $(\zeta_i , -1)= L-M$ (longitude - meridian) basis is given by $S T^{-1} |\s\rangle$ where 

\[
{S} =   \left|
 \begin{array}{ccc}
  1/2 & \sqrt{2}/2 & 1/2 \\
  \sqrt{2}/2  & 0 & - \sqrt{2}/2  \\
  1/2 & - \sqrt{2}/2  & 1/2
 \end{array}
\right|
\tn{ and }
{T} =   \left|
 \begin{array}{ccc}
  1 & 0 & 0 \\
  0 & e^{i \pi/8} & 0  \\
  0 & 0  & -1
 \end{array}
\right|
\]
in the $\{1, \s, \e\}$ basis.  We check that $S T^{-1} \,\,| \s\rangle = e^{-i \pi/8}\sqrt{2}/2 \,\, |1\rangle - e^{-i \pi/8} \sqrt{2}/2 \,\,|\e\rangle$.  This calculation will soon be justified below.

\vskip.2in \epsfxsize=3.75in \centerline{\epsfbox{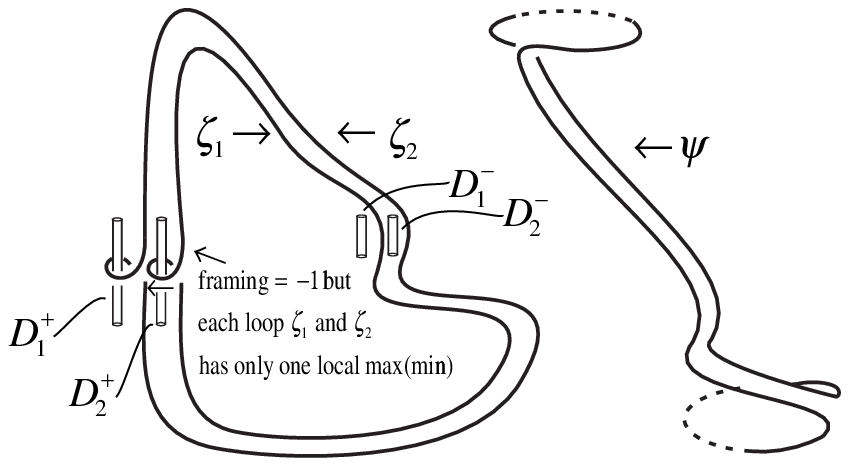}.}
{\centerline{Figure 18}}

Ignoring the measurements $\xi_1 \ldots \xi_n$ which create the
$|1\rangle-$labeled passage between the inner punctures of $P$ as in Figure 14, we must execute three titled measurements along $\psi$, $ \xi_1$, and
$\xi_2$.  We have just shown that in all cases the outcomes for
charge $(\xi_1)$ and charge $(\xi_2)$ are independent and either $1$ or $\psi$.  We previously verified charge $(\psi)=1$ or $\e$.  From this it will follow that
the protocol produces 
$ {g_1} = \left|
 \begin{smallmatrix}
  1 & 0 \\
  0 & \pm e^{\pi i/4}  
 \end{smallmatrix}
\right| $ iff charge $(\psi) = 1$  and $\left|
\begin{smallmatrix}
  \pm e^{\pi i/4} & 0  \\
               0 & 1
 \end{smallmatrix}
\right| $ iff  charge $(\psi)
= \psi$, where $+$ occurs if charge $(\zeta_1) \cdot$ charge $(\zeta_2)=1$ and $-$ if charge $(\zeta_1)\cdot$ charge $(\zeta_2) = \e$. (Our notation is motivated by fusion
rules:  $1 \otimes 1=1$, $ 1 \otimes\psi = \psi \otimes 1 =
\psi$, and $\psi \otimes \psi =1$.)  In all eight measurement outcomes we have, up to an overall phase, implemented either $g_1, g_1^{3}, g_1^{5}$ or $g_1^{7}$ within the cyclic group of order $=8$, $Z/8Z$ generated by $g_1$.  Thus our protocol generates some random walk on $Z/8Z$ with one of four possible steps
determined by fair coins.  Since we know the measurement outcomes, we see where we are walking and 
may iterate the protocol until we arrive at $g_1$.  Again this is
efficient.

As the reader has probably anticipated, in terms of [BK], measuring
charge $(\psi)$, $\psi$ as in Figure 18, corresponds to measuring the
charge on $\psi$ of Figure 11.  Measuring the charges on $\zeta_1$ and
$\zeta_2$ correspond to the double Dehn twist in a manner which we now
explain.

In section 2 we commented that projection to charge sectors on a
loop $\gamma$ does not become well-defined (or in fact the eigenspaces
themselves) until $\gamma$ has a normal framing.  If the
physical Hilbert space $V(T)$ for a torus $T$ is $V=$ span $\{1,
\sigma, \psi\}$ in the meridinal basis $L$ if we wish to
transform to the framing $=k$ basis, $L+kM =$ longitude
$+k$(meridian), we must compute as follows (See [Wa].):(cuff, seem){\footnote{To write $V(T^2)\cong {\underset{\tn{particle types,} a}{\Sigma}} V_{a \overline{a}}(S^1 \times I)$, we need to select a circle, the $\lq\lq$cuff'', to cut the torus along and a dual circle, the $\lq\lq$seam'', to trivialize the resulting annulus as $S^1 \times I$.}}
$= (M,L)
\stackrel{\textnormal{twist}{^k}}{\longrightarrow}(M,L+kM)\stackrel{S}{\longrightarrow}
(L+kM,M) $ the composition given by:
$
S T^k
$.

The curves $\zeta_1$ and $\zeta_2$ have been described as having framing $=-1$
this means the tip of the frame vector links $-1$ with its base as
it moves around the loop and that $k$, above is $-1$.  It is a fundamental identity of the
$\lq\lq$Kirby calculus'' that $-1-$framed surgery on a simple linking
circle imparts a $+1-$Dehn twist.


\vskip.2in \epsfxsize=4.5in \centerline{\epsfbox{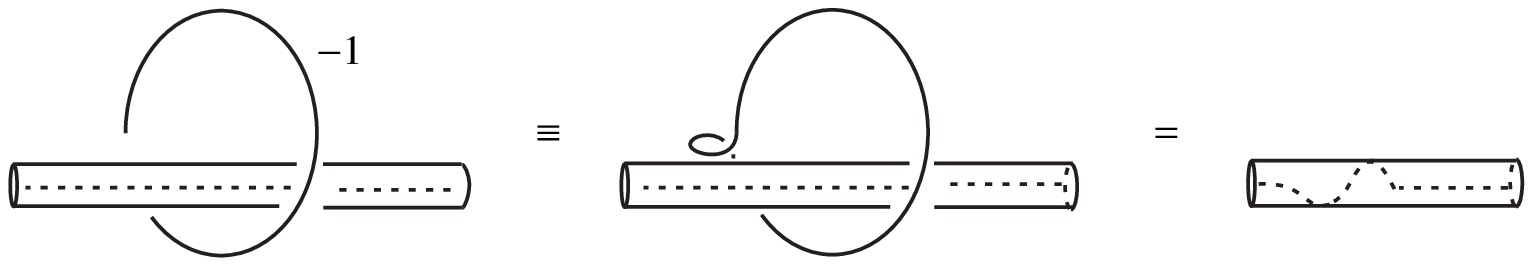}}
{\centerline{Figure 19}}



The meaning of $\lq\lq$surgery'' is that a tubular neighborhood of
the loop is deleted and then glued back so that the meridian disk is glued to 
the circle defined by the tip of the frame vector.  Obviously we can
neither twist nor surger Gallium Arsenide, but if we measure the
particle content of a (framed!) loop $\gamma$ in the interior of a
$2+1-$dimensional space-time, and the result is $1$, we have (up to an
overall normalization factor, corresponding to capping a $2-$sphere) accomplished surgery on $\gamma$ as far as Chern-Simons theory
is concerned.  Similarly if we measure a nontrivial particle
$\sigma$ or $\psi$ we have still done a kind of surgery but now
the reglued solid torus has a particle $\lq\lq$Polyakov loop'' ($\sigma$ or $\psi$
resp.) running along its core. This is $\sigma$  or $\psi = Z$ (solid torus, Polyakov loop) $\in V(T^2)$ expressed in meridinal basis.

From the $S-$matrices, we know that (w.r.t. the labeling
in Figure 11) charge $(\alpha) = 1 - \psi$ iff charge $(\psi)
\cdot$ charge $(\delta) = 1$  and charge $(\alpha)
= \sigma$ iff charge $(\psi) \cdot$ charge $(\delta) = \psi$.  Thus the nontrivial phase arises in the upper left or lower right entry of our gate-matrix according to whether charge $(\psi)=1$ or $\e$.

In translating between Figures 11 and 18, $\alpha$ corresponds to
untwisted copies of the $\zeta$'s, $(\zeta_1, 0)$ and $(\zeta_2, 0)$.

Measuring $(\zeta_1, -1)$ and $(\zeta_2 , -1)$ results in a squared Dehn
twist around $\alpha$ with two Polyakov loops appearing,
labeled by some particle type $1$ or $\psi$, (but not $\s$!) parallel
to $\alpha$, say at $\alpha \times 1/3$ and $\alpha \times 2/3$ in a
product structure.

The Polyakov loops cannot carry $\s$ since $S T^{-1} \,\,| \s \rangle$ has no  $| \s\rangle$ component.  Because two $\e$'s must fuse to $1$, only the total charge, charge $(\zeta_1 , -1)\cdot$ charge $(\zeta_2 , -1) =1$ or $\e$ is relevant to the action of the gate.  There are two cases: when charge $(\delta) \cdot$ charge $(\psi)=1$, then charge $(\a)=1-\e$, and the effect of an $\e-$ Polyakov loop can be localized as that of a $\e-$core, circle$\times 1/2$ in annulus$\times[0,1]$, where the boundaries of the annulus are labeled by $|1\rangle$ (or $|\e\rangle$).  With either labeling, the $\e-$ Polyakov loop contributes no additional phase.  In contrast, in the second case when charge $(\delta)\cdot$ charge$(\psi)=\e$ and charge $(\a)=\s$, the localized model is an $\e-$ Polyakov loop at level $1/2$ in annulus $\times \,\,[0,1]$ with all four boundaries labeled by $\s$.  In this case the Polyakov loop contributes a phase$-1$. 

\vskip.2in \epsfxsize=4.5in \centerline{\epsfbox{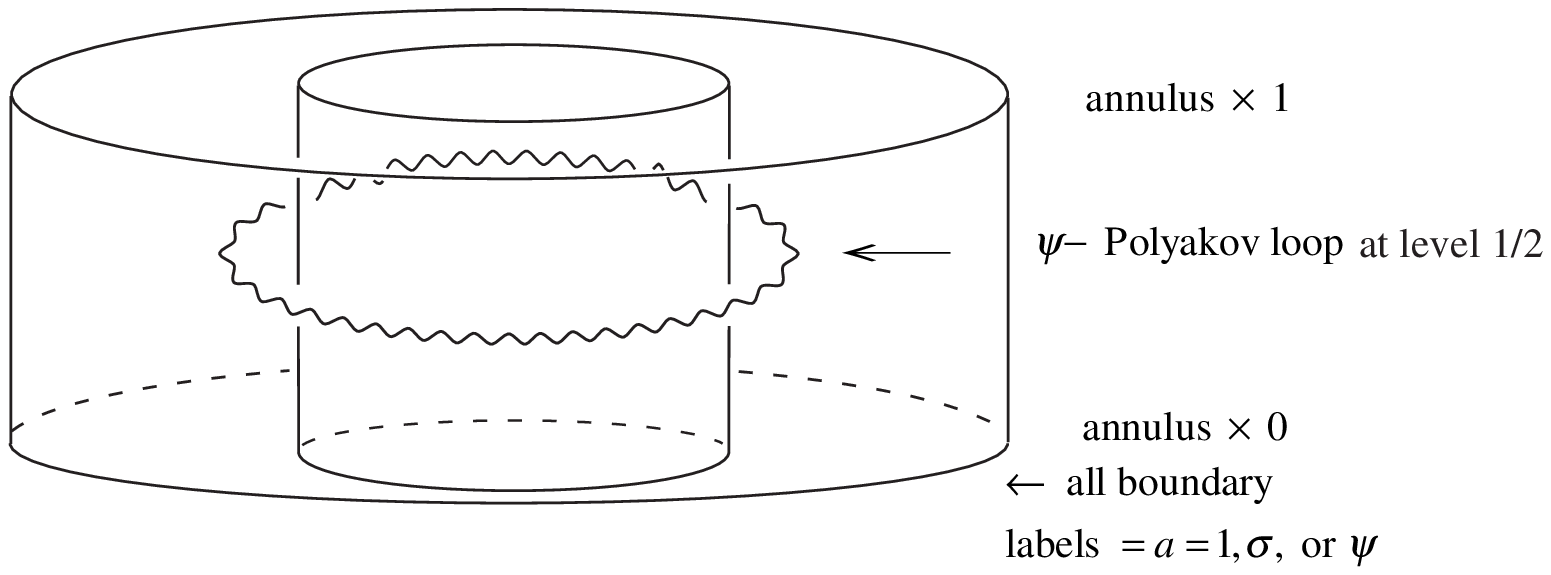}.}
{\centerline{Figure 20}}

The phase which the $\e-$Polyakov loop adds to the identity (product) morphism is:

\vskip.01in \epsfxsize=4.75in \centerline{\epsfbox{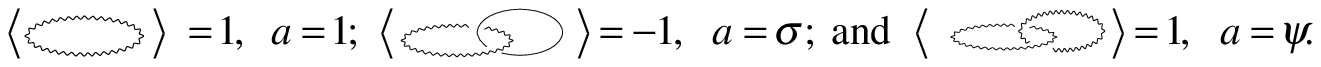}}

This completes the verification that our protocol implements $g_1, g_1^{3}, g_1^{5}$ or 
$g_1^{7}$ in the cases as follows: 
\begin{align*}
\tn{charge} (\psi) &=1, \tn{charge} (\zeta , -1)\cdot \tn{charge} (\zeta_2 , -1)=1 \Longrightarrow g_1 \\
\tn{charge} (\psi)& =\e, \tn{charge} (\zeta , -1)\cdot \tn{charge}(\zeta_2 , -1)=\e ,  \Longrightarrow g_{1}^{3} \\
\tn{charge} (\psi) &=1, \tn{charge} (\zeta , -1)\cdot \tn{charge} (\zeta_2 , -1)=\e , \Longrightarrow g_{1}^{5}\\
\tn{charge} (\psi)& =\e, \tn{charge} (\zeta , -1)\cdot \tn{charge} (\zeta , -1)=1 \Longrightarrow g_{1}^{7}
\end{align*}

{\bf\centerline{Appendix A: The effect of adding or deleting $1-$handles to pants $\times I$.}}

The time history of adding and then breaking a band between the inner boundary components $\beta, \omega$ or a twice punctured disk $P$ is topologically the addition of a $1-$ handle $(D^1 \times D^2, \partial D^1 \times D^2)$ to $P \times I$; call the result $W=P\times I \cup 1 -$handle.

\vskip.2in \epsfxsize=4.5in \centerline{\epsfbox{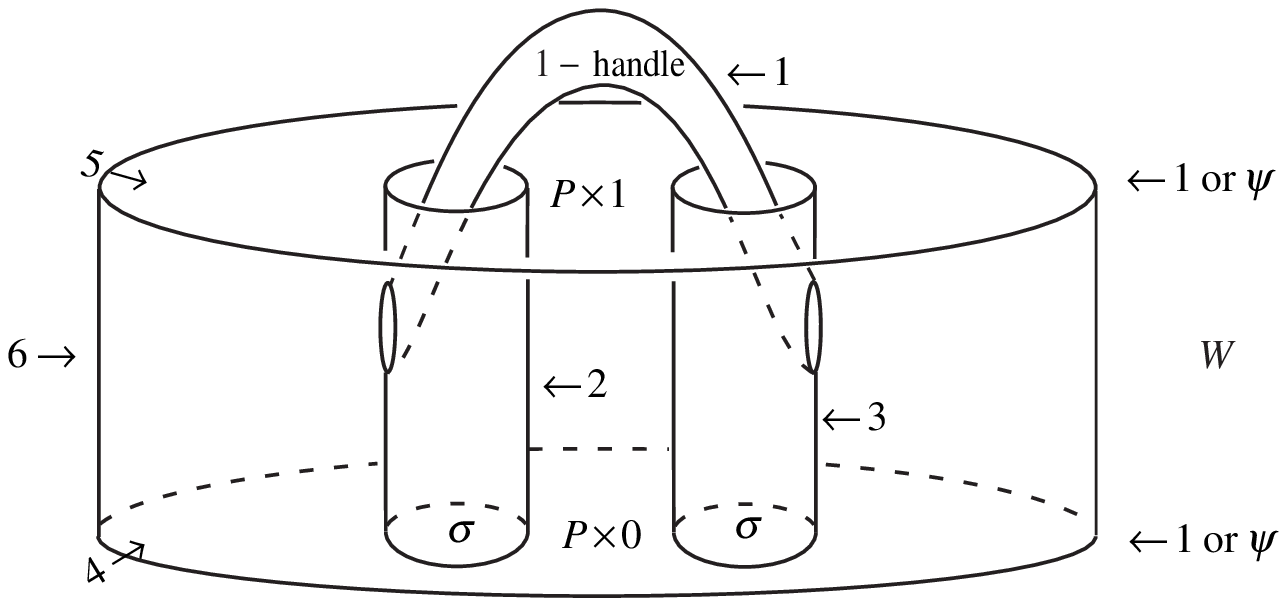}}
{\centerline{Figure 21}}
Warning:  $W$ is drawn with time $\neq Z-$coordinate since the add/break procedure for the band $B$ does not imbed in $(2+1)-$dimensions.

While it is axiomatic that products correspond in a TQFT to identity morphisms, it is a small calculation that $W$ induces the identity (rather than say a phase gate) on the qubit supported on $P$.  The general principle is that if a surface which bounds a $3-$manifold is broken up into sub-surfaces by (particle)labeled loops, then the $3-$manifold canonically specifies a vector in the tensor product of the (relative) physical Hilbert spaces.  Letting $x=1$ or $\e$ for the outer label, $W$ specifies a vector $\psi_1$ in:
$V_{0,0} \otimes V_{0,\s,\s} \otimes V_{0,\s,\s}\otimes V_{\s,\s, x} \otimes  V_{\s,\s,x}^{\ast} \otimes V_{x,x}$, where the factors come from subsurfaces $1, \ldots, 6$ in Figure 22.  The zero label in the first three factors is dictated by the presence of the disks in $W$ capping the boundary of the first component (a cylinder).  The gluing axion [W] or [T] tells us that removing the $1-$handle determines a canonical isomorphism to $Z(P\times I)$ carrying $\psi_1$ to $\psi_0$ in  $V_{0}^{\ast} \otimes V_{0} \otimes V_{0,\s,\s}\otimes V_{0, \s,\s} \otimes  V_{\s,\s,x} \otimes V_{\s,\s, x}^{\ast} \otimes V_{x,x}$.  After supplying the canonical base vectors $\beta_0^\ast \in V_0^\ast , \beta_{0,\s,\s} \in V_{0,\s, \s}$ and $\beta_{x, x}\in V_{x,x}, \psi_1$ is canonically identified with id $\in$ Hom $(V_{\s,\s,x}) \cong V_{\s,\s, x}^{\ast} \otimes V_{\s,\s,x}$.  Note that no $x-$dependent phase has entered the calculation.  Thus we have proved, in the abstract language of TQFTs, that adding and then breaking a band induces the identity operator on the qubit supported in $P$.

The situation is rather different if, instead we cut out a band to join the internal punctures and then restore it.  (In other language fuse the internal punctures and then separate them.)  We will even assume that we can use the electric charge of the $\s$, which we take to be $+\frac{e}{4}$ on both punctures to ensure that is energetically favorable (and hence necessary) that when we split the previously fused puncture back in two, each resulting puncture again carries a $+\frac{e}{4}$ charged $\s$.  Our calculation will show that even in this situation we have not acted on the $P-$ qubit via the identity by rather a POVM $\a \cong \left|
 \begin{smallmatrix}
  1 & 0 \\
  0 & 1  
 \end{smallmatrix}
\right| + \beta \cong
\left|
 \begin{smallmatrix}
  1 & 0 \\
  0 & -1  
 \end{smallmatrix}
\right|$ where $\a (\beta)$ is the probability for $\gamma$ in Figure 23 to carry charge $1(\e)$.  The moral, in general, is that operations which add quantum media (in this case $5/2-$ FQHE fluid) are reversible - simply delete what was previously added, whereas operations which delete are often irreversible.

\vskip.2in \epsfxsize=4.5in \centerline{\epsfbox{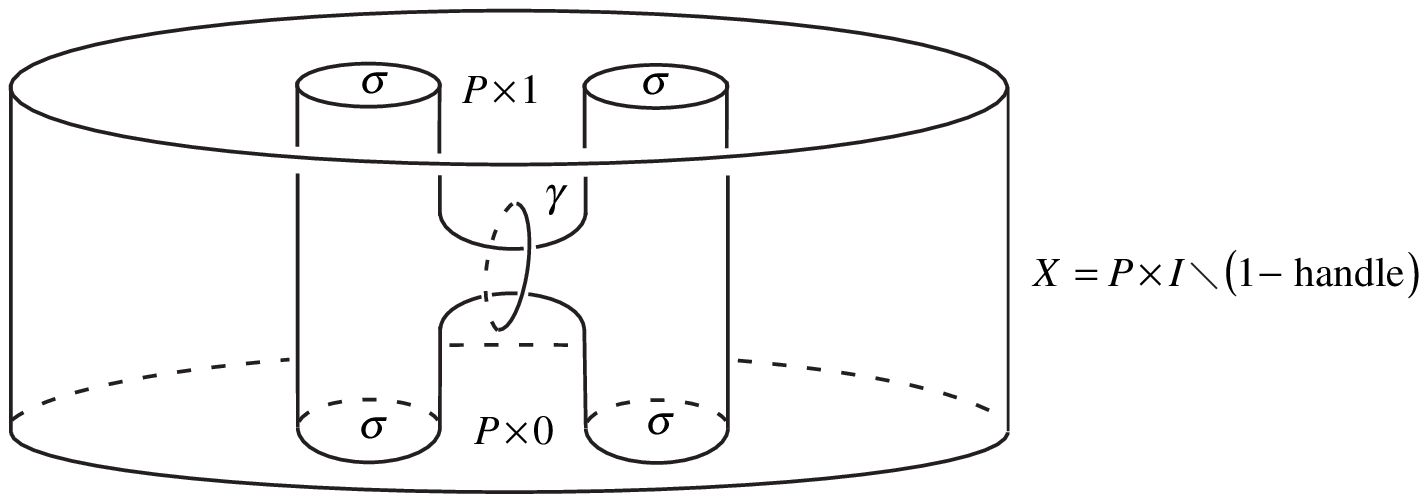}}
{\centerline{Figure 22}}

In particle flight (Feynmann diagram) notation Figure 22 is either Figure 23a or figure 23b, according to the charge of $\gamma$.

\vskip.2in \epsfxsize=3in \centerline{\epsfbox{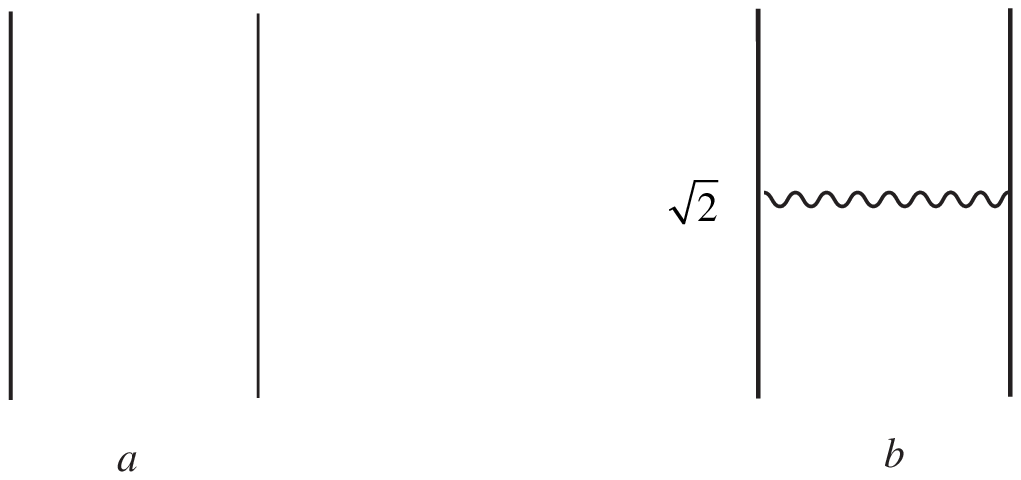}}
{\centerline{Figure 23}}

Figure 22a certainly represents the identity acting on the $P-$qubit.  The operator given by Figure 22b is clearly diagonal in the $(1, \e)$ basis.  To compute the phases of the diagonal entries we pair with the orthonormal exterior basis \includegraphics[width=1.75cm, height=.25in]{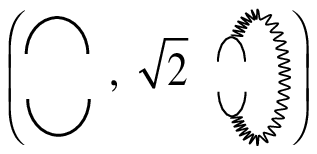} and use the Kauffman relations to extract expectation values. In case $a$ we get:
\vskip.2in \epsfxsize=3in \centerline{\epsfbox{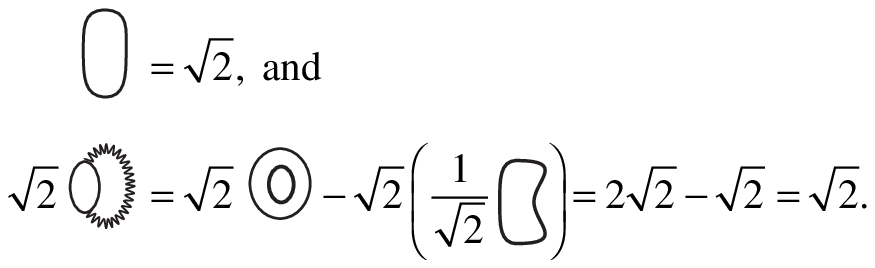}}

In this case $b$ we get:
\vskip.2in \epsfxsize=3.75in \centerline{\epsfbox{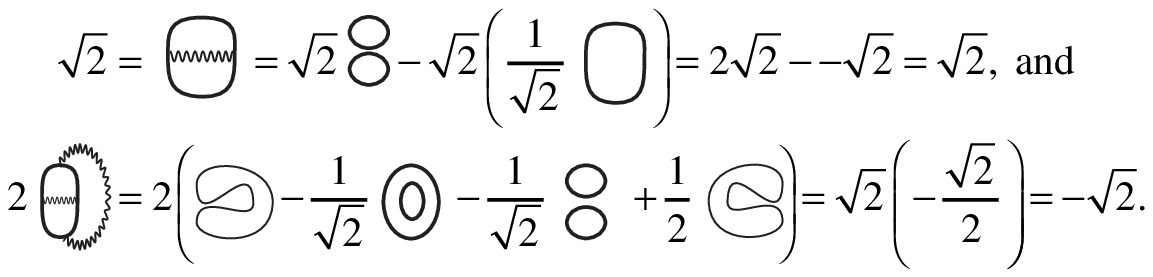}}
The strange $\sqrt{2}$ factor is actually $S_{0 0} = S_{\e\e}$ which has entered because we have not rescaled the dual physical Hilbert space by $1/S_{xx}$ prior to gluing.  Taking this axiomatic factor into account (see [BK] or [W]) we obtain the claimed formula.

\end{document}